\documentclass[aip,reprint]{revtex4-1}
\usepackage{graphicx}
\usepackage{dcolumn}
\usepackage{bm}

\begin{document}

\title{Wiedemann-Franz Relation and Thermal-transistor Effect in Suspended Graphene}

\author{S. Yi\u{g}en and A.~R. Champagne}
\email{a.champagne@concordia.ca}
\affiliation{Department of Physics, Concordia University, Montr\'{e}al, Qu\'{e}bec, H4B 1R6 Canada}

\date{\today}

\begin{abstract}
We extract experimentally the electronic thermal conductivity, $K_{e}$, in suspended graphene which we dope using a back-gate electrode. We make use of two-point dc electron transport at low bias voltages and intermediate temperatures (50 - 160 K), where the electron and lattice temperatures are decoupled. The thermal conductivity is proportional to the charge conductivity times the temperature, confirming that the Wiedemann-Franz relation is obeyed in suspended graphene. We extract an estimate of the Lorenz coefficient as 1.1 to 1.7 $\times 10^{-8}$ W $\Omega$K$^{-2}$. $K_e$ shows a transistor effect and can be tuned with the back-gate by more than a factor of 2 as the charge carrier density ranges from $\approx$ 0.5 to 1.8 $\times 10^{11}$cm$^{-2}$.
\end{abstract}

\keywords{Graphene, Thermal conductivity, Wiedemann-Franz, Thermal transistor, electron-phonon}

\maketitle

Graphene's electronic thermal conductivity, $K_{e}$, describes how easily Dirac charge carriers (electron and hole quasiparticles) can carry energy. In low-disorder graphene at moderate temperatures ($<$ 200 - 300 K), the energy transfer rate between charge carriers and acoustic phonons is extremely slow \cite{Gabor11,Song11,Viljas11,Fong12,DasSarma13, Yigen13}. Thus, $K_{e}$ impacts how a hot electron cools down, and the efficiency of charge harvesting in graphene optoelectronic devices \cite{Song11,Gabor11, Tielrooij13}. Moreover, understanding and controlling $K_{e}$ could help develop graphene bolometers capable of detecting single terahertz photons \cite{Fong12, Fong13}. There are theoretical calculations of $K_{e}$ \cite{Saito07, Muller08, Foster09}, and recent experimental data near the charge neutrality point (CNP) in clean suspended graphene \cite{Yigen13} and in disordered samples at very low temperatures \cite{Fong12,Fong13}. However, a detailed mapping of $K_{e}$  vs charge density at intermediate temperatures is lacking. Understanding how $K_{e}$ in clean (suspended) graphene depends on charge density, $n$, and the electronic temperature, $T_{e}$, is crucial for applications. An important fundamental question is whether the Wiedemann-Franz (WF) law, $K_{e} = \sigma LT_{e}$ where $\sigma$ is the charge conductivity, and $L$ is the Lorenz number, is obeyed in graphene. In clean graphene at low charge densities (hydrodynamic regime), strong electron-electron interactions could lead to departures from the generalized WF law \cite{Muller08, Foster09}.

\begin{figure}
\includegraphics [width=3.25in]{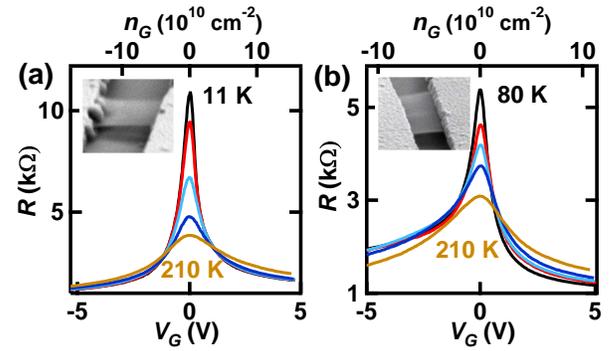}
\caption{\label{}(Color online.) Suspended and annealed graphene devices. (a) $R$ of Sample A vs. gate voltage, $V_G$, at $T_{e} = T$ = 11, 50, 100, 150 and 210 K, and $V_{B} =$ 0.5 mV. (b) $R$ - $V_G$ data for Sample B at $T_{e} = T$ = 80, 100, 125, 150 and 210 K, and $V_{B} =$ 5 mV. The insets show tilted SEM images of the suspended graphene transistors: Sample A (650 nm long), and a device identical to Sample B (400 nm long). For clarity, the data curves in (a) and (b) are slightly shifted along the $V_{G}$ axis so that all the maxima line up at $V_{G}=$ 0, the shifts in (a) range from -0.3 to -0.45 V, and in (b) from 0 to 0.2 V.}
\end{figure}

We report $K_{e}$ in monolayer graphene extracted from carefully calibrated dc electron transport measurements following a method we previously discussed \cite{Yigen13}. We study a temperature range of $T=$ 50 - 160 K, where the electron and lattice temperatures are very well decoupled in low-disorder graphene \cite{Gabor11,Song11,Viljas11,Fong12,DasSarma13, Yigen13}, over a charge density range of $\approx$ 0.5 to 1.8 $\times 10^{11}$cm$^{-2}$. We extract data in the hole and electron doped regimes from two high-mobility suspended devices. The extracted $K_e$ are compared with predictions from the WF law. The agreement between the WF relation and measurements is very good for both devices over the $n$ range studied and $T$ up to 160 K. The value of $L$ is $\approx$ 0.5 - 0.7 $L_{o}$, where $L_{o}$ is the Lorenz factor for metals. We observe a sudden jump in the extracted thermal conductivity above 160 K which is consistent with the onset of strong coupling between electrons and acoustic phonons \cite{DasSarma13}. Finally, we observe a thermal transistor effect consistent with the WF prediction, where $K_{e}$ can be tuned by more than a factor of 2 with a back-gate voltage, $V_{G}$, ranging up to $\pm$ 5 V. Throughout the text we use $T$ to designate the lattice (cryostat) temperature, and $T_{e}$ for the average electron temperature in the suspended devices. At very low bias, $|V_{B}| \leq$ 1 mV, $T = T_{e}$.

Figure 1 (a)-(b) shows dc two-point resistance data, $R = V_{B}/I$, versus gate voltage, $V_{G}$, which controls the charge density, $n_{G}$ (upper axis) for Samples A and B respectively. From the width of the $R$ maximum at low $T$, we extract a half-width-half-maximum, HWHM, of 0.45 and 0.95 V for Samples A and B. These HWHMs correspond to an impurity induced charge density of $n^{*} \approx$ 1.5 and 2.1 $\times 10^{10} $ cm$^{-2}$. For clarity, the data in Fig.\ 1 is slightly shifted along the $V_{G}$ axis so that all the maxima (Dirac points) line up at $V_{G}=$ 0, the shifts in panel (a) range from -0.3 to -0.45 V at various $T$, and in panel (b) from 0 to 0.2 V. The insets in Fig.\ 1 show scanning electron microscope (SEM) tilted images of Sample A and a sample identical to Sample B.

We confirmed, using optical contrast and Raman spectroscopy, that both samples are single-layer graphene. Sample A is 650 nm long, 675 nm wide, and suspended 140 $\pm$ 10 nm above the substrate (AFM measurement) which consists of 100 $\pm$ 2 nm of SiO$_2$ (ellipsometry measurement) on degenerately-doped Si which is used as a back-gate electrode. Sample B is 400 nm long, 0.97 $\mu$m wide, and suspended 227 $\pm$ 10 nm above a 74 $\pm$ 2 nm SiO$_2$ film on Si. To prepare the samples, we used exfoliated graphene, and standard e-beam lithography to define Ti(5nm)/Au(80nm) contacts. The samples were suspended with a wet BOE etch such that their only thermal connection was to the gold contacts. We annealed the devices using Joule heating \textit{in situ} by flowing a large current in the devices\cite{Bolotin08, Yigen13} (see Supplemental Information (SI) section 1). Annealing and all subsequent measurements were done under high vacuum, $\leq 10^{-6}$ Torr.
\begin{figure}
\includegraphics[width=3.25in]{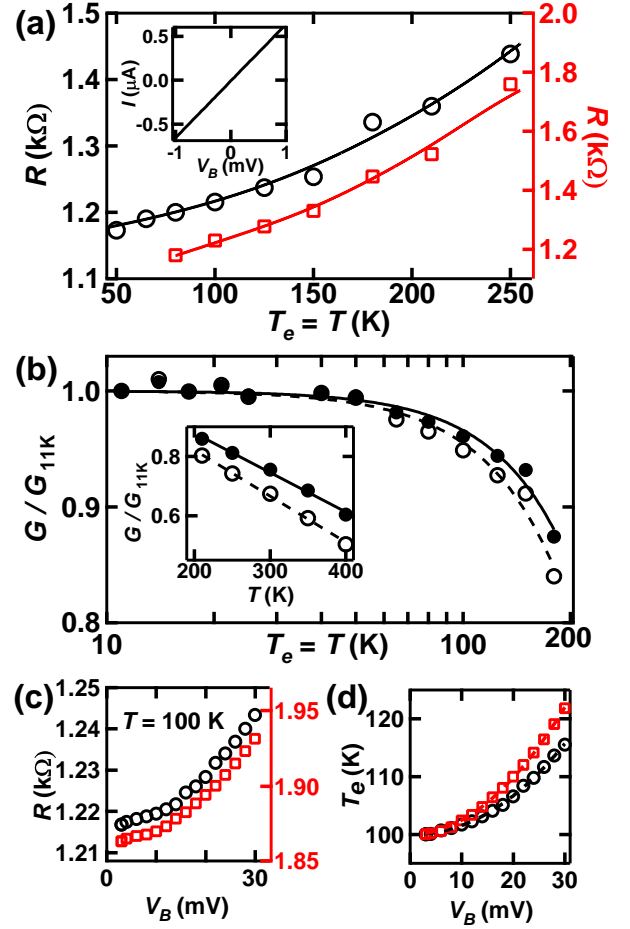}
\caption{\label{}(Color online.) Electron thermometry and electron Joule heating. (a) $R$ vs $T_{e}$ in Sample A (circles, left axis) and Sample B (squares, right axis), respectively hole ($n_{G} =$ -1.8$\times 10^{11}$ cm$^{-2}$) and electron doped ($n_{G} =$ 1.1$\times 10^{11}$ cm$^{-2}$). The solid lines are numerically interpolated curves used for thermometry. Inset: example of $I-V_{B}$ data at 100 K for Sample A whose slope is used to extract $R$, $|V_{B}| < 1$ mV such that $T_{e} = T$. (b) Relative conductance $G/G_{11 K}$ of Sample A vs $T_{e}  = T$. The solid circles show the raw two-point data, and the open circles the data using $R_c =$ 239 $\Omega$. The solid and dashed lines are power law fits consistent with charge impurity scattering. Inset: $G/G_{11 K}$ of Sample A at high $T$ showing a linear decrease in $R$ consistent with acoustic phonon scattering. (c) $R$ vs $V_B$ data for Sample A at $T =$ 100 K, and $n_{G} =$ -1.8 (circles, left axis) and -0.8 (squares, right axis) $\times 10^{11}$ cm$^{-2}$ (i.e. $V_{G} =$ -5.3 and -2.3 V). Joule heating due to $V_B$ raises the flake's average $T_e$ above $T$. (d) $T_e$ vs $V_{B}$ in Sample A extracted from (c) using the calibration in (a) and similar curves.}
\end{figure}

To minimize contact resistance, $R_{c}$, the devices were fabricated with large contact areas between the gold electrodes and graphene crystals, ranging from 1.1 to 3 $\mu m^2$ per contact. An upper bound for series resistance, $R_{o}$, which includes both the contact resistance, $R_{c}$, and the resistance from neutral scatterers, can be extracted from the two-point $R-V_{G}$ curves\cite{Dean10} in Fig. 1 (see SI section 2). The extracted series resistances for Sample A are $R_{o-h} \approx$ 477 $\pm$ 53 and $R_{o-e} \approx$ 944 $\pm$ 80 $\Omega$ for hole and electron doping, respectively. The difference between $R_{o-h}$ and $R_{o-e}$ is understood as an additional $p-n$ barrier for the electron due to doping from the gold electrodes \cite{Castro10}. For Sample B, we find $R_{o-h} =$ 1563 $\Omega$ and $R_{o-e} =$ 812 $\Omega$. We note that series resistance is smallest for hole doping in Sample A and for electron doping in Sample B. In annealed samples, oxygen desorbs from the gold contacts and changes the work function of the electrodes. This means that graphene under the gold electrodes can be either electron doped or hole doped \cite{} depending on the thoroughness of the contact annealing\cite{Castro10,Heinze02,Giovannetti08}. To minimize the effect of $R_{c}$ on our data, we study the lowest resistance side of the Dirac point for each Sample. This allows us to study hole transport in Sample A and electron transport in Sample B. Since $R_{o}$ includes both $R_{c}$ and the resistance due to neutral scatterers in the channel, we conservatively set $R_{c}$ = $R_{o}/2$ with an uncertainty ranging up to $R_{c-max}=R_{o}$, and down to $R_{c-min} =$ lowest reported resistance for Au/graphene\cite{Ifuku13} with similar $n$, which is $\approx$ 100 $\Omega . \mu$m$^{2}$. Thus, in the following data analysis we use for Samples A and B, $R_{c-A}=$ 239 $\pm ^{239}_{120}$ and $R_{c-B}=$ 406 $\pm ^{406}_{281}$. We extract a conservative estimate of the charge carrier mobility in our devices, over the $n$ and $T_{e}$ range studied, as $\mu=\sigma/(n_{tot}e) \approx$ 3.5 $\times$ 10 $^{4}$ cm$^{2}$/V.s, where $n_{tot}$ is the total carrier density including the gate, impurity and thermal doping \cite{Dorgan10,Yigen13} (SI section 3). Based on the reported thermal conductance of Au/Ti/Graphene and Graphene/SiO$_{2}$ interfaces\cite{Koh10}, the thermal resistance of the contacts can safely be neglected\cite{Dorgan13} compared to our $K_{e}$ data presented below.

Figure 2 summarizes our approach to extract $K_{e}$ in suspended high-mobility graphene, whose details we previously discussed \cite{Yigen13}. We repeat some of the discussion of our methods because the charge densities studied here are much higher than in Ref. 6, which leads to several important changes. Figure 2(a)-(b) presents how we monitor the charged quasiparticle temperature in our devices by monitoring $R$, and Fig.\ 2(c)-(d) shows how we can controllably heat-up these quasiparticles at a temperature slightly above the contacts' temperature via Joule heating. By combining these two capabilities and using the heat equation, we will later extract $K_{e}$ vs $T$ and $n$.

Figure 2(a) shows the two-point dc $R$ vs cryostat temperature, $T$, calibration curves for Sample A (circles, left axis), and Sample B (squares, right axis) which are respectively hole-doped with a gate-induced density of $n_{G} =$ -1.8$\times 10^{11}$ cm$^{-2}$ and electron-doped with $n_{G} =$ 1.1$\times 10^{11}$ cm$^{-2}$. $R = V_{B}/I$ is extracted from the slope of the $I-V_{B}$ data\cite{Rnote} as shown in the inset of Fig.\ 2(a). Note that for $\pm$ 1 mV bias no Joule heating effect is present and $T_{e} = T$. The $T_{e}$ dependence of the data shows a metallic behavior with $R$ increasing with $T_{e}$. The interpolated lines in Fig.\ 2(a), and similar curves, will be used as secondary thermometry curves to monitor $T_e$ in the devices.

Figure 2(b) shows the relative conductance $G(T)/G_{11 K}$ for Sample A extracted from Fig.\ 2(a) and similar data. The solid circles show the raw two-point data, and the open circles the data after subtracting $R_c = R_{o}/2$. The $T$ dependence of $G = 1/R$ in graphene, at modest charge density, is strongly dependent on the type of charge transport. We fit the data in Fig. 2(b) with a function $G/G_{11 K} = 1 - AT^{p}$, and extract $p$ = 2.1 $\pm$ 0.2 for both curves. This $T$-dependence strongly supports diffusive charge transport dominated by long-range charge impurities, as reported in previous experiments on high-mobility devices \cite{Bolotin08, Du08, DasSarma11} and expected theoretically \cite{DasSarma13}. The inset of Fig.\ 2(b) shows that $G/G_{11 K}$ of Sample A decreases linearly for $T \geq$ 200 K, which suggests a relatively strong acoustic phonon scattering above this temperature, as expected theoretically \cite{DasSarma13}. Sample B shows a qualitatively identical behavior of its $R$ vs $T$ in Fig. 2(a), but the absence of low temperature data proscribes an accurate fit of its dependence. We will focus our measurements on the $T <$ 200 K range, where both samples are in the diffusive regime (SI section 3) with scattering predominantly due to charged impurities. This scattering is elastic, and its $T_{e}$ dependence (used for thermometry) comes mostly from the temperature dependence of its screening \cite{DasSarma11}.

Electron-electron scattering between charge carriers is inelastic. By applying a $V_{B}$ one can inject high-energy carriers in the suspended device which then thermalize with the carriers in the sample and raise $T_{e}$ in the suspended graphene relative to the temperature in the gold contacts. Note that when writing $T_{e}$, we always refer to the \textit{average} temperature of charged quasiparticles in our devices. We demonstrate controlled Joule self-heating of the electrons to apply a thermal bias $\Delta T = T_{e} - T$ between the suspended graphene and the electrodes (cryostat temperature). Figure 2(c) shows $R$ vs $V_{B}$ for Sample A at $T = $ 100 K and $n_{G} =$ -1.8 $\times$ 10$^{11}$ (circles, left axis) and -0.8 $\times 10^{11}$ cm$^{-2}$ (squares, right axis). Sample B data is shown in SI sec.\ 4. $R$ increases monotonically with increasing $V_{B}$, at all $T$. We restrict our measurements to $V_{B}$ $\leq 27$ meV. We have previously argued\cite{Yigen13} that in our high-mobility devices, under such low $V_{B}$ and in the $T$ range we study, the change in $R$ is caused by Joule heating of the charge carriers\cite{Viljas11,Yigen13}. Using the curves, $R$ vs $T_{e}$ and $R$ vs $V_{B}$, we extract $T_{e}$ vs $V_{B}$ as shown for Sample A in Fig.\ 2(d). We fit a power law (dashed lines) $T_e = 100 + BV_{B}^{x}$, and find $x=$ 1.93 $\pm$ 0.04 for both data sets, as expected for Joule heating where $T_e \propto V_{B}^{2}$. Figures 3(d) and S3(b) show that the accuracy with which $T_{e}$ can be extracted is much better than 1 K. The smooth dependence of $T_{e}$ on $V_{B}$ at all $T$ is consistent with electrons having a well defined temperature as predicted by calculations of the $e-e$ collision length\cite{Li13} (see SI of Ref. 6), and confirmed by the $K_{e}$ data shown below.
\begin{figure}
\includegraphics[width=3.25in]{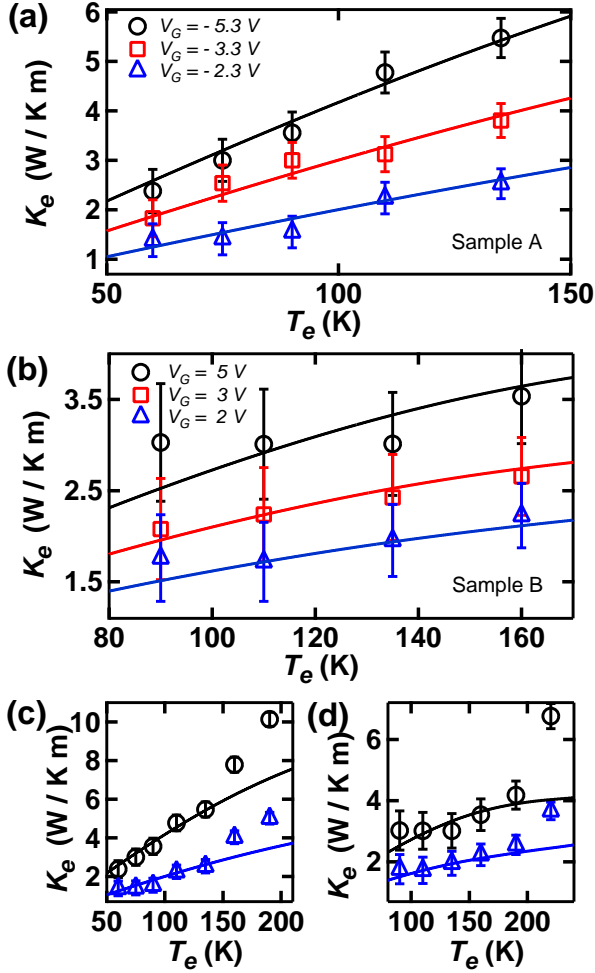}
\caption{\label{}(Color online.) Wiedemann-Franz (WF) law. (a) The electronic thermal conductivity, $K_{e}$, of Sample A in the hole-doped regime vs $T_{e}$ for $\Delta T = T_{e} - T =$ 10 K. The circle, square and triangle data show $K_{e}$ at $V_{G} =$ -5.3, -3.3 and -2.3 V respectively corresponding to $n_{tot, T=0}\approx$ -1.8, -1.1, -0.8 $\times 10^{11}$ cm$^{-2}$. The solid lines are given by the WF relation $K_{WF} = L\sigma T_{e}$ with $L =$ 0.45, 0.53 and 0.55 $\times$ $L_{o}$ respectively. (b) $K_{e}$ vs $T_{e}$ for $\Delta T =$ 10 K for Sample B in the electron-doped regime. The circle, square and triangle data show $K_{e}$ at $V_{G} =$ 5, 3 and 2 V corresponding to $n_{tot, T=0}\approx$ 1.1, 0.7, 0.5 $\times 10^{11}$ cm$^{-2}$. The solid lines are the $K_{WF}$ with $L=$ 0.66, 0.68, 0.7 $\times$ $L_{o}$. (c) and (d) show the same data as in (a) and (b) up to higher $T_{e}$ where the apparant departure between the data and WF prediction is understood as the onset of electron-phonon coupling.}
\end{figure}

We use a 1-d heat equation\cite{Yigen13} to extract $K_{e}$ in our devices, and find $K_{e} = \frac{Q\ell^{2}}{12\Delta T}$, where $ \Delta T = T_{e} - T$, $Q = RI^{2}/W\ell h$ is the Joule heating power per unit volume, $W$ the width, $\ell$ the length, $h$ = 0.335 nm the thickness, and $T_{e}$ is the electronic temperature averaged over $\ell$. In Fig.\ 3(a)-(b) we plot $K_{e}$ vs $T_{e}$ for Samples A and B for $\Delta T$ = 10 K, where the circle, square and triangle data show $K_{e}$ at $n_{tot, T=0}\approx$ -1.8, -1.1, -0.8 $\times 10^{11}$ cm$^{-2}$ for Sample A, and $n_{tot, T=0}\approx$ 1.1, 0.7, 0.5 $\times 10^{11}$ cm$^{-2}$ for Sample B. The quantity $n_{tot, T=0}$ refers to the total charge density induced by $V_{G}$ and charged impurities (SI section 3). We clearly observe that $K_{e}$ increases with both $n$ and $T_{e}$ in both samples. For instance, $K_{e}$ ranges from roughly 1 W/K.m at 60 K and $n=$ -8 $\times$ 10$^{10}$cm$^{-2}$ to 5 W/K.m at 135 K and $n=$ -1.8 $\times$ 10$^{11}$cm$^{-2}$ for Sample A. Error bars representing the uncertainty on the extracted $K_{e}$ are shown in Fig. 3 (SI section 5). We confirmed that the $V_{B}$ needed to create $\Delta T$ did not dope significantly the samples or affect the measured $K_{e}$ \cite{VBnote}. The thermoelectric voltages in our devices are negligible compared to $V_{B}$\cite{Zuev09, Hwang09}.

We test the WF law in our samples, which have a mobility of $\mu \approx$ 3.5 $\times$ 10 $^{4}$ cm$^{2}$/V.s, as a function of $T_{e}$ and $n$. While the Lorenz number in most metals is close to $L_{o} =$ 2.44 $\times$ 10$^{-8}$ $W \Omega K^{-2}$, it is well known that its value can be reduced in semiconductors at low charge density \cite{Bian07, Minnich09}. The solid lines in Fig.\ 3 show $K_{e-WF}$ given by the WF law using the measured $\sigma$ and extracted $T_{e}$ (Fig.\ 2), with $L$ used as the single fitting parameter. The WF relation holds for both Samples at all $T_{e}$ between 50 K and 160 K, and densities $n_{h}=$ -1.8 to -0.8 $\times 10^{11}$ cm$^{-2}$ and $n_{e}=$ 0.5 to 1.1 $\times 10^{11}$ cm$^{-2}$. For Sample A, $L =$ 0.45, 0.53 and 0.55 $\times$ $L_{o}$, and for Sample B $L=$ 0.66, 0.68, 0.7 $\times$ $L_{o}$ (triangle, square, and circle data, respectively). The main uncertainty on $L$ comes from the uncertainty on $R_{c}$, and corresponds to $\pm ^{0.1}_{0.2}$ $L_{o}$ for Sample A, and $\pm$ 0.4 $L_{o}$ for Sample B. We note that the qualitative temperature and density dependence of the data in Fig. 3, and the agreement with the WF law, is preserved even if we use either the maximum $R_{c} = R_{o}$ or minimum $R_{c} =$ 120 $\Omega$ (SI section 6). The increase in $L$ as $n$ increases is consistent with previous studies in semiconductors where the value of $L$ tends toward $L_{o}$ at higher carrier density \cite{Minnich09}.
\begin{figure}
\includegraphics[width=3.25in]{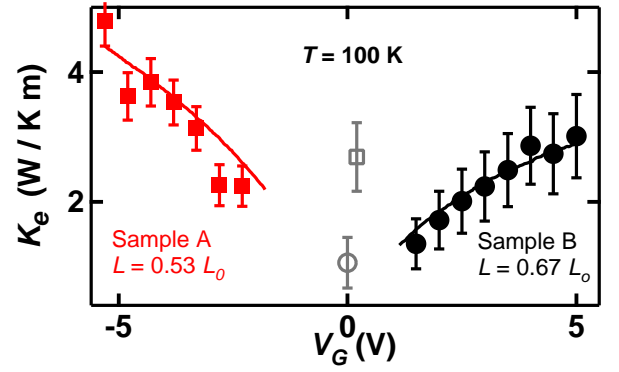}
\caption{\label{}(Color online.) Electronic thermal conductivity transistor effect. $K_{e}$ vs $V_{G}$ data for Samples A (solid squares, $n <$ 0) and B (solid circles, $n >$ 0), for $T$ = 100 K and $\Delta T =$ 10 K. $V_{G} \approx 0$ data (open symbols, $\Delta T =$ 5 K) are from Ref. 6. The solid lines show $K_{WF}$ predicted by the Wiedemann-Franz relation using a Lorenz number of $L$ = 0.53 $L_{o}$ and 0.67 $L_{o}$ for A and B respectively.}
\end{figure}

Electron to acoustic phonon coupling is very weak in clean graphene at moderate doping and temperature ($<$ 200 - 300 K) \cite{Gabor11,Song11,Viljas11,Yigen13, DasSarma13}, but increases at higher $n$ and $T$. In the context of our experiment, if the thermal energy conductance between electrons and phonons $G_{e-p}$ is non-negligeable compared to the electronic thermal conductance $G_{e}$, the heat conductivity we extract is a mixture of $K_{e}$ and $K_{e-p}$ in parallel. As can be seen in Fig. 3(c)-(d), above $\approx$ 160 K the extracted $K$ no longer agrees with the WF prediction (solid line), indicating that we cannot isolate $K_{e}$ for $T_{e} > 160 K$. Previously we found that we could extract $K_{e}$ up to 300 K in samples whose $n$ was very close to the CNP \cite{Yigen13}, suggesting that $e-p$ coupling is weaker at lower $n$ as expected theoretically\cite{DasSarma13}. In Fig. 3(c)-(d), the departure between the $K$ data and WF prediction starts around 150 K for Samples A and 200 K for Sample B. The different $T$ ranges over which $K_{e}$ dominates in the two devices comes from the ratio $G_{e}/G_{e-ph} =W/(W\ell) = 1/\ell$ which is 60 $\%$ larger for Sample B than Sample A.

Figure 4 shows the extracted $K_{e}$ vs $V_{G}$ at $T_{e} = T + \Delta T = 100 + 10$ K for Sample A (B) as solid red squares (black circles). For reference only, we also show two data points (open grey symbols) close to $V_{G}\approx$ 0 which are taken from Ref.\ 6 for the same Samples. We cannot extract $K_{e}$ at intermediate $n$, i.e. when 0.3 V $\leq |V_{G}|\leq $ 1.5 V. This is because while $R$ smoothly increases with $T_{e}$ in the metal-like regime (Fig.\ 2(a)), $|V_{G}|\geq $ 1.5 V, and smoothly decrease in the insulator-like regime\cite{Yigen13}, $|V_{G}|\leq $ 0.3 V, the $R$ vs $T_{e}$ behavior does not act as a good thermometer at intermediate densities. We also note that the WF relation we discuss in this work, $K_{e} =\sigma L T_{e}$, only applies in the degenerately doped regime where the Fermi energy $\mu >> k_{B}T_{e}$, and we focus our discussion on this regime. The solid symbol data in Fig.\ 4 show that $K_{e}$ is tuned by the charge carrier density in the samples. The solid lines are the WF values $K_{e-WF}$ calculated using the measured $\sigma$, $T_{e} =$ 110 K, and setting $L$ = 0.53 and 0.67 $L_{o}$ for Samples A and B. The agreement between the WF law and data in the doped regime is excellent. Even using a modest $V_{G}$ range, $K_{e}$ could be tuned by a factor of $\geq$ 2 in Fig.\ 4. This is a very strong thermal-transistor effect (with the caveat that $K_{p} >> K_{e}$ \cite{Balandin11,Pop12}). This could have applications in optoelectronics. A larger $K_{e}$ means that when a charge carrier is excited by a photon, it can travel a larger distance and excite additional carriers before it thermalizes with the lattice. Thus, more of the photon energy is harvested as electrical current \cite{Tielrooij13}. Additionally, a tunable $K_{e}$ implies a tunable $C_{e}$ which could be used to optimize bolometric applications of graphene \cite{Fong12, Fong13}.

In summary, we fabricated high quality suspended graphene devices. We used a self-thermometry and self-heating method \cite{Yigen13} to extract the electronic thermal conductivity in doped graphene. We report for the first time $K_e$ in suspended graphene over a broad range of $T$ and $n$. The data presented clearly demonstrates that $K_e \propto \sigma T$, which confirms that the Wiedemann-Franz law holds in high-mobility ($\mu \approx$ 3.5 $\times$ 10 $^{4}$ cm$^{2}$/V.s) suspended graphene over our accessible temperature range, 50 K - 160 K. This temperature range is limited at high-temperature by a turning on of the electron-phonon coupling, which prevents us from isolating $K_{e}$ at higher $T$. The clear onsets of the electron-phonon coupling (Fig. 2(b), and 3(c)-(d)) between 150 K- 200K is consistent with theoretical calculations \cite{DasSarma13}. We studied charge densities of holes and electrons ranging up to 1.8 $\times$ 10$^{11}$ cm$^{-2}$ and found Lorenz numbers $L\approx$ 0.5 - 0.7 $L_{o}$, where $L_{o}$ is the standard Lorenz number for metals. The quality of the agreement between the data and the WF relation in Figs.\ 3 and 4 is not affected by the uncertainty on the extracted Lorenz numbers (SI section 6). Finally, we demonstrated a strong thermal-transistor effect where we could tune $K_{e}$ by more than a factor of 2 by applying only a few volts to a gate electrode.

In the future, these measurements could be extended to even cleaner devices at lower densities to study possible corrections to the generalized WF relation due to strong electron-electron interactions\cite{Muller08, Foster09}. The demonstrated density control of $K_e$ could be useful to make energy harvesting optoelectronic devices \cite{Song11,Gabor11,Tielrooij13} and sensitive bolometers\cite{Fong12,Fong13}. We thank Vahid Tayari, James Porter and Andrew McRae for technical help and discussions. This work was supported by NSERC, CFI, FQRNT, and Concordia University. We made use of the QNI cleanrooms.


\begin{thebibliography}{31}%
\makeatletter
\providecommand \@ifxundefined [1]{%
 \@ifx{#1\undefined}
}%
\providecommand \@ifnum [1]{%
 \ifnum #1\expandafter \@firstoftwo
 \else \expandafter \@secondoftwo
 \fi
}%
\providecommand \@ifx [1]{%
 \ifx #1\expandafter \@firstoftwo
 \else \expandafter \@secondoftwo
 \fi
}%
\providecommand \natexlab [1]{#1}%
\providecommand \enquote  [1]{``#1''}%
\providecommand \bibnamefont  [1]{#1}%
\providecommand \bibfnamefont [1]{#1}%
\providecommand \citenamefont [1]{#1}%
\providecommand \href@noop [0]{\@secondoftwo}%
\providecommand \href [0]{\begingroup \@sanitize@url \@href}%
\providecommand \@href[1]{\@@startlink{#1}\@@href}%
\providecommand \@@href[1]{\endgroup#1\@@endlink}%
\providecommand \@sanitize@url [0]{\catcode `\\12\catcode `\$12\catcode
  `\&12\catcode `\#12\catcode `\^12\catcode `\_12\catcode `\%12\relax}%
\providecommand \@@startlink[1]{}%
\providecommand \@@endlink[0]{}%
\providecommand \url  [0]{\begingroup\@sanitize@url \@url }%
\providecommand \@url [1]{\endgroup\@href {#1}{\urlprefix }}%
\providecommand \urlprefix  [0]{URL }%
\providecommand \Eprint [0]{\href }%
\providecommand \doibase [0]{http://dx.doi.org/}%
\providecommand \selectlanguage [0]{\@gobble}%
\providecommand \bibinfo  [0]{\@secondoftwo}%
\providecommand \bibfield  [0]{\@secondoftwo}%
\providecommand \translation [1]{[#1]}%
\providecommand \BibitemOpen [0]{}%
\providecommand \bibitemStop [0]{}%
\providecommand \bibitemNoStop [0]{.\EOS\space}%
\providecommand \EOS [0]{\spacefactor3000\relax}%
\providecommand \BibitemShut  [1]{\csname bibitem#1\endcsname}%
\let\auto@bib@innerbib\@empty
\bibitem [{\citenamefont {Gabor}\ \emph {et~al.}(2011)\citenamefont {Gabor},
  \citenamefont {Song}, \citenamefont {Ma}, \citenamefont {Nair}, \citenamefont
  {Taychatanapat}, \citenamefont {Watanabe}, \citenamefont {Taniguchi},
  \citenamefont {Levitov},\ and\ \citenamefont {Jarillo-Herrero}}]{Gabor11}%
  \BibitemOpen
  \bibfield  {author} {\bibinfo {author} {\bibfnamefont {N.~M.}\ \bibnamefont
  {Gabor}}, \bibinfo {author} {\bibfnamefont {J.~C.~W.}\ \bibnamefont {Song}},
  \bibinfo {author} {\bibfnamefont {Q.}~\bibnamefont {Ma}}, \bibinfo {author}
  {\bibfnamefont {N.~L.}\ \bibnamefont {Nair}}, \bibinfo {author}
  {\bibfnamefont {T.}~\bibnamefont {Taychatanapat}}, \bibinfo {author}
  {\bibfnamefont {K.}~\bibnamefont {Watanabe}}, \bibinfo {author}
  {\bibfnamefont {T.}~\bibnamefont {Taniguchi}}, \bibinfo {author}
  {\bibfnamefont {L.~S.}\ \bibnamefont {Levitov}}, \ and\ \bibinfo {author}
  {\bibfnamefont {P.}~\bibnamefont {Jarillo-Herrero}},\ }\href@noop {}
  {\bibfield  {journal} {\bibinfo  {journal} {Science}\ }\textbf {\bibinfo
  {volume} {334}},\ \bibinfo {pages} {648} (\bibinfo {year}
  {2011})}\BibitemShut {NoStop}%
\bibitem [{\citenamefont {Song}\ \emph {et~al.}(2011)\citenamefont {Song},
  \citenamefont {Rudner}, \citenamefont {Marcus},\ and\ \citenamefont
  {Levitov}}]{Song11}%
  \BibitemOpen
  \bibfield  {author} {\bibinfo {author} {\bibfnamefont {J.~C.~W.}\
  \bibnamefont {Song}}, \bibinfo {author} {\bibfnamefont {M.~S.}\ \bibnamefont
  {Rudner}}, \bibinfo {author} {\bibfnamefont {C.~M.}\ \bibnamefont {Marcus}},
  \ and\ \bibinfo {author} {\bibfnamefont {L.~S.}\ \bibnamefont {Levitov}},\
  }\href@noop {} {\bibfield  {journal} {\bibinfo  {journal} {Nano Lett.}\
  }\textbf {\bibinfo {volume} {11}},\ \bibinfo {pages} {4688} (\bibinfo {year}
  {2011})}\BibitemShut {NoStop}%
\bibitem [{\citenamefont {Viljas}\ \emph {et~al.}(2011)\citenamefont {Viljas},
  \citenamefont {Fay}, \citenamefont {Wiesner},\ and\ \citenamefont
  {Hakonen}}]{Viljas11}%
  \BibitemOpen
  \bibfield  {author} {\bibinfo {author} {\bibfnamefont {J.~K.}\ \bibnamefont
  {Viljas}}, \bibinfo {author} {\bibfnamefont {A.}~\bibnamefont {Fay}},
  \bibinfo {author} {\bibfnamefont {M.}~\bibnamefont {Wiesner}}, \ and\
  \bibinfo {author} {\bibfnamefont {P.~J.}\ \bibnamefont {Hakonen}},\
  }\href@noop {} {\bibfield  {journal} {\bibinfo  {journal} {Phys. Rev. B}\
  }\textbf {\bibinfo {volume} {83}},\ \bibinfo {pages} {205421} (\bibinfo
  {year} {2011})}\BibitemShut {NoStop}%
\bibitem [{\citenamefont {Fong}\ and\ \citenamefont {Schwab}(2012)}]{Fong12}%
  \BibitemOpen
  \bibfield  {author} {\bibinfo {author} {\bibfnamefont {K.~C.}\ \bibnamefont
  {Fong}}\ and\ \bibinfo {author} {\bibfnamefont {K.~C.}\ \bibnamefont
  {Schwab}},\ }\href@noop {} {\bibfield  {journal} {\bibinfo  {journal} {Phys.
  Rev. X}\ }\textbf {\bibinfo {volume} {2}},\ \bibinfo {pages} {031006}
  (\bibinfo {year} {2012})}\BibitemShut {NoStop}%
\bibitem [{\citenamefont {Das~Sarma}\ and\ \citenamefont
  {Hwang}(2013)}]{DasSarma13}%
  \BibitemOpen
  \bibfield  {author} {\bibinfo {author} {\bibfnamefont {S.}~\bibnamefont
  {Das~Sarma}}\ and\ \bibinfo {author} {\bibfnamefont {E.~H.}\ \bibnamefont
  {Hwang}},\ }\href@noop {} {\bibfield  {journal} {\bibinfo  {journal} {Phys.
  Rev. B}\ }\textbf {\bibinfo {volume} {87}},\ \bibinfo {pages} {035415}
  (\bibinfo {year} {2013})}\BibitemShut {NoStop}%
\bibitem [{\citenamefont {Yigen}\ \emph {et~al.}(2013)\citenamefont {Yigen},
  \citenamefont {Tayari}, \citenamefont {Island}, \citenamefont {Porter},\ and\
  \citenamefont {Champagne}}]{Yigen13}%
  \BibitemOpen
  \bibfield  {author} {\bibinfo {author} {\bibfnamefont {S.}~\bibnamefont
  {Yigen}}, \bibinfo {author} {\bibfnamefont {V.}~\bibnamefont {Tayari}},
  \bibinfo {author} {\bibfnamefont {J.~O.}\ \bibnamefont {Island}}, \bibinfo
  {author} {\bibfnamefont {J.~M.}\ \bibnamefont {Porter}}, \ and\ \bibinfo
  {author} {\bibfnamefont {A.~R.}\ \bibnamefont {Champagne}},\ }\href@noop {}
  {\bibfield  {journal} {\bibinfo  {journal} {Phys. Rev. B}\ }\textbf {\bibinfo
  {volume} {87}},\ \bibinfo {pages} {241411} (\bibinfo {year}
  {2013})}\BibitemShut {NoStop}%
\bibitem [{\citenamefont {Tielrooij}\ \emph {et~al.}(2013)\citenamefont
  {Tielrooij}, \citenamefont {Song}, \citenamefont {Jensen}, \citenamefont
  {Centeno}, \citenamefont {Pesquera}, \citenamefont {Elorza}, \citenamefont
  {Bonn}, \citenamefont {Levitov},\ and\ \citenamefont
  {Koppens}}]{Tielrooij13}%
  \BibitemOpen
  \bibfield  {author} {\bibinfo {author} {\bibfnamefont {K.~J.}\ \bibnamefont
  {Tielrooij}}, \bibinfo {author} {\bibfnamefont {J.~C.~W.}\ \bibnamefont
  {Song}}, \bibinfo {author} {\bibfnamefont {S.~A.}\ \bibnamefont {Jensen}},
  \bibinfo {author} {\bibfnamefont {A.}~\bibnamefont {Centeno}}, \bibinfo
  {author} {\bibfnamefont {A.}~\bibnamefont {Pesquera}}, \bibinfo {author}
  {\bibfnamefont {A.~Z.}\ \bibnamefont {Elorza}}, \bibinfo {author}
  {\bibfnamefont {M.}~\bibnamefont {Bonn}}, \bibinfo {author} {\bibfnamefont
  {L.~S.}\ \bibnamefont {Levitov}}, \ and\ \bibinfo {author} {\bibfnamefont
  {F.~H.~L.}\ \bibnamefont {Koppens}},\ }\href@noop {} {\bibfield  {journal}
  {\bibinfo  {journal} {Nature Physics}\ }\textbf {\bibinfo {volume} {9}},\
  \bibinfo {pages} {248--252} (\bibinfo {year} {2013})}\BibitemShut {NoStop}%
\bibitem [{\citenamefont {Fong}\ \emph {et~al.}(2013)\citenamefont {Fong},
  \citenamefont {Wollman}, \citenamefont {Ravi}, \citenamefont {Chen},
  \citenamefont {Clerk}, \citenamefont {Shaw}, \citenamefont {Leduc},\ and\
  \citenamefont {Schwab}}]{Fong13}%
  \BibitemOpen
  \bibfield  {author} {\bibinfo {author} {\bibfnamefont {K.~C.}\ \bibnamefont
  {Fong}}, \bibinfo {author} {\bibfnamefont {E.}~\bibnamefont {Wollman}},
  \bibinfo {author} {\bibfnamefont {R.}~\bibnamefont {Ravi}}, \bibinfo {author}
  {\bibfnamefont {W.}~\bibnamefont {Chen}}, \bibinfo {author} {\bibfnamefont
  {A.~A.}\ \bibnamefont {Clerk}}, \bibinfo {author} {\bibfnamefont {M.~D.}\
  \bibnamefont {Shaw}}, \bibinfo {author} {\bibfnamefont {H.~G.}\ \bibnamefont
  {Leduc}}, \ and\ \bibinfo {author} {\bibfnamefont {K.~C.}\ \bibnamefont
  {Schwab}},\ }\href@noop {} {\bibfield  {journal} {\bibinfo  {journal} {arXiv:
  1308.2265}\ } (\bibinfo {year} {2013})}\BibitemShut {NoStop}%
\bibitem [{\citenamefont {Saito}, \citenamefont {Nakamura},\ and\ \citenamefont
  {Natori}(2007)}]{Saito07}%
  \BibitemOpen
  \bibfield  {author} {\bibinfo {author} {\bibfnamefont {K.}~\bibnamefont
  {Saito}}, \bibinfo {author} {\bibfnamefont {J.}~\bibnamefont {Nakamura}}, \
  and\ \bibinfo {author} {\bibfnamefont {A.}~\bibnamefont {Natori}},\
  }\href@noop {} {\bibfield  {journal} {\bibinfo  {journal} {Phys. Rev. B}\
  }\textbf {\bibinfo {volume} {76}},\ \bibinfo {pages} {115409} (\bibinfo
  {year} {2007})}\BibitemShut {NoStop}%
\bibitem [{\citenamefont {Muller}, \citenamefont {Fritz},\ and\ \citenamefont
  {Sachdev}(2008)}]{Muller08}%
  \BibitemOpen
  \bibfield  {author} {\bibinfo {author} {\bibfnamefont {M.}~\bibnamefont
  {Muller}}, \bibinfo {author} {\bibfnamefont {L.}~\bibnamefont {Fritz}}, \
  and\ \bibinfo {author} {\bibfnamefont {S.}~\bibnamefont {Sachdev}},\
  }\href@noop {} {\bibfield  {journal} {\bibinfo  {journal} {Phys. Rev. B}\
  }\textbf {\bibinfo {volume} {78}},\ \bibinfo {pages} {115406} (\bibinfo
  {year} {2008})}\BibitemShut {NoStop}%
\bibitem [{\citenamefont {Foster}\ and\ \citenamefont
  {Aleiner}(2009)}]{Foster09}%
  \BibitemOpen
  \bibfield  {author} {\bibinfo {author} {\bibfnamefont {M.~S.}\ \bibnamefont
  {Foster}}\ and\ \bibinfo {author} {\bibfnamefont {I.~L.}\ \bibnamefont
  {Aleiner}},\ }\href@noop {} {\bibfield  {journal} {\bibinfo  {journal} {Phys.
  Rev. B}\ }\textbf {\bibinfo {volume} {79}},\ \bibinfo {pages} {085415}
  (\bibinfo {year} {2009})}\BibitemShut {NoStop}%
\bibitem [{\citenamefont {Bolotin}\ \emph {et~al.}(2008)\citenamefont
  {Bolotin}, \citenamefont {Sikes}, \citenamefont {Jiang}, \citenamefont
  {Klima}, \citenamefont {Fudenberg}, \citenamefont {Hone}, \citenamefont
  {Kim},\ and\ \citenamefont {Stormer}}]{Bolotin08}%
  \BibitemOpen
  \bibfield  {author} {\bibinfo {author} {\bibfnamefont {K.~I.}\ \bibnamefont
  {Bolotin}}, \bibinfo {author} {\bibfnamefont {K.~J.}\ \bibnamefont {Sikes}},
  \bibinfo {author} {\bibfnamefont {Z.}~\bibnamefont {Jiang}}, \bibinfo
  {author} {\bibfnamefont {M.}~\bibnamefont {Klima}}, \bibinfo {author}
  {\bibfnamefont {G.}~\bibnamefont {Fudenberg}}, \bibinfo {author}
  {\bibfnamefont {J.}~\bibnamefont {Hone}}, \bibinfo {author} {\bibfnamefont
  {P.}~\bibnamefont {Kim}}, \ and\ \bibinfo {author} {\bibfnamefont {H.~L.}\
  \bibnamefont {Stormer}},\ }\href@noop {} {\bibfield  {journal} {\bibinfo
  {journal} {Solid State Comm.}\ }\textbf {\bibinfo {volume} {146}},\ \bibinfo
  {pages} {351} (\bibinfo {year} {2008})}\BibitemShut {NoStop}%
\bibitem [{\citenamefont {Dean}\ \emph {et~al.}(2010)\citenamefont {Dean},
  \citenamefont {Young}, \citenamefont {Meric}, \citenamefont {Lee},
  \citenamefont {Wang}, \citenamefont {Sorgenfrei}, \citenamefont {Watanabe},
  \citenamefont {Taniguchi}, \citenamefont {Kim}, \citenamefont {Shepard},\
  and\ \citenamefont {Hone}}]{Dean10}%
  \BibitemOpen
  \bibfield  {author} {\bibinfo {author} {\bibfnamefont {C.~R.}\ \bibnamefont
  {Dean}}, \bibinfo {author} {\bibfnamefont {A.~F.}\ \bibnamefont {Young}},
  \bibinfo {author} {\bibfnamefont {I.}~\bibnamefont {Meric}}, \bibinfo
  {author} {\bibfnamefont {C.}~\bibnamefont {Lee}}, \bibinfo {author}
  {\bibfnamefont {L.}~\bibnamefont {Wang}}, \bibinfo {author} {\bibfnamefont
  {S.}~\bibnamefont {Sorgenfrei}}, \bibinfo {author} {\bibfnamefont
  {K.}~\bibnamefont {Watanabe}}, \bibinfo {author} {\bibfnamefont
  {T.}~\bibnamefont {Taniguchi}}, \bibinfo {author} {\bibfnamefont
  {P.}~\bibnamefont {Kim}}, \bibinfo {author} {\bibfnamefont {K.~L.}\
  \bibnamefont {Shepard}}, \ and\ \bibinfo {author} {\bibfnamefont
  {J.}~\bibnamefont {Hone}},\ }\href@noop {} {\bibfield  {journal} {\bibinfo
  {journal} {Nature Nanotechnol.}\ }\textbf {\bibinfo {volume} {5}},\ \bibinfo
  {pages} {722} (\bibinfo {year} {2010})}\BibitemShut {NoStop}%
\bibitem [{\citenamefont {Castro}\ \emph {et~al.}(2010)\citenamefont {Castro},
  \citenamefont {Ochoa}, \citenamefont {Katsnelson}, \citenamefont {Gorbachev},
  \citenamefont {Elias}, \citenamefont {Novoselov}, \citenamefont {Geim},\ and\
  \citenamefont {Guinea}}]{Castro10}%
  \BibitemOpen
  \bibfield  {author} {\bibinfo {author} {\bibfnamefont {E.~V.}\ \bibnamefont
  {Castro}}, \bibinfo {author} {\bibfnamefont {H.}~\bibnamefont {Ochoa}},
  \bibinfo {author} {\bibfnamefont {M.~I.}\ \bibnamefont {Katsnelson}},
  \bibinfo {author} {\bibfnamefont {R.~V.}\ \bibnamefont {Gorbachev}}, \bibinfo
  {author} {\bibfnamefont {D.~C.}\ \bibnamefont {Elias}}, \bibinfo {author}
  {\bibfnamefont {K.~S.}\ \bibnamefont {Novoselov}}, \bibinfo {author}
  {\bibfnamefont {A.~K.}\ \bibnamefont {Geim}}, \ and\ \bibinfo {author}
  {\bibfnamefont {F.}~\bibnamefont {Guinea}},\ }\href@noop {} {\bibfield
  {journal} {\bibinfo  {journal} {Phys. Rev. Lett.}\ }\textbf {\bibinfo
  {volume} {105}},\ \bibinfo {pages} {266601} (\bibinfo {year}
  {2010})}\BibitemShut {NoStop}%
\bibitem [{\citenamefont {Heinze}\ \emph {et~al.}(2002)\citenamefont {Heinze},
  \citenamefont {Tersoff}, \citenamefont {Martel}, \citenamefont {Derycke},
  \citenamefont {Appenzeller},\ and\ \citenamefont {Avouris}}]{Heinze02}%
  \BibitemOpen
  \bibfield  {author} {\bibinfo {author} {\bibfnamefont {S.}~\bibnamefont
  {Heinze}}, \bibinfo {author} {\bibfnamefont {J.}~\bibnamefont {Tersoff}},
  \bibinfo {author} {\bibfnamefont {R.}~\bibnamefont {Martel}}, \bibinfo
  {author} {\bibfnamefont {V.}~\bibnamefont {Derycke}}, \bibinfo {author}
  {\bibfnamefont {J.}~\bibnamefont {Appenzeller}}, \ and\ \bibinfo {author}
  {\bibfnamefont {P.}~\bibnamefont {Avouris}},\ }\href@noop {} {\bibfield
  {journal} {\bibinfo  {journal} {Phys. Rev. Lett.}\ }\textbf {\bibinfo
  {volume} {89}},\ \bibinfo {pages} {106801} (\bibinfo {year}
  {2002})}\BibitemShut {NoStop}%
\bibitem [{\citenamefont {Giovannetti}\ \emph {et~al.}(2008)\citenamefont
  {Giovannetti}, \citenamefont {Khomyakov}, \citenamefont {Brocks},
  \citenamefont {Karpan}, \citenamefont {van~den Brink},\ and\ \citenamefont
  {Kelly}}]{Giovannetti08}%
  \BibitemOpen
  \bibfield  {author} {\bibinfo {author} {\bibfnamefont {G.}~\bibnamefont
  {Giovannetti}}, \bibinfo {author} {\bibfnamefont {P.~A.}\ \bibnamefont
  {Khomyakov}}, \bibinfo {author} {\bibfnamefont {G.}~\bibnamefont {Brocks}},
  \bibinfo {author} {\bibfnamefont {V.~M.}\ \bibnamefont {Karpan}}, \bibinfo
  {author} {\bibfnamefont {J.}~\bibnamefont {van~den Brink}}, \ and\ \bibinfo
  {author} {\bibfnamefont {P.~J.}\ \bibnamefont {Kelly}},\ }\href@noop {}
  {\bibfield  {journal} {\bibinfo  {journal} {Phys. Rev. Lett.}\ }\textbf
  {\bibinfo {volume} {101}},\ \bibinfo {pages} {026803} (\bibinfo {year}
  {2008})}\BibitemShut {NoStop}%
\bibitem [{\citenamefont {Ifuku}\ \emph {et~al.}(2013)\citenamefont {Ifuku},
  \citenamefont {Nagashio}, \citenamefont {Nishimura},\ and\ \citenamefont
  {Toriumi}}]{Ifuku13}%
  \BibitemOpen
  \bibfield  {author} {\bibinfo {author} {\bibfnamefont {R.}~\bibnamefont
  {Ifuku}}, \bibinfo {author} {\bibfnamefont {K.}~\bibnamefont {Nagashio}},
  \bibinfo {author} {\bibfnamefont {T.}~\bibnamefont {Nishimura}}, \ and\
  \bibinfo {author} {\bibfnamefont {A.}~\bibnamefont {Toriumi}},\ }\href@noop
  {} {\bibfield  {journal} {\bibinfo  {journal} {arXiv: 1307.0690}\ } (\bibinfo
  {year} {2013})}\BibitemShut {NoStop}%
\bibitem [{\citenamefont {Dorgan}, \citenamefont {Bae},\ and\ \citenamefont
  {Pop}(2010)}]{Dorgan10}%
  \BibitemOpen
  \bibfield  {author} {\bibinfo {author} {\bibfnamefont {V.~E.}\ \bibnamefont
  {Dorgan}}, \bibinfo {author} {\bibfnamefont {M.~H.}\ \bibnamefont {Bae}}, \
  and\ \bibinfo {author} {\bibfnamefont {E.}~\bibnamefont {Pop}},\ }\href@noop
  {} {\bibfield  {journal} {\bibinfo  {journal} {Appl. Phys. Lett.}\ }\textbf
  {\bibinfo {volume} {97}},\ \bibinfo {pages} {082112} (\bibinfo {year}
  {2010})}\BibitemShut {NoStop}%
\bibitem [{\citenamefont {Koh}\ \emph {et~al.}(2010)\citenamefont {Koh},
  \citenamefont {Bae}, \citenamefont {Cahill},\ and\ \citenamefont
  {Pop}}]{Koh10}%
  \BibitemOpen
  \bibfield  {author} {\bibinfo {author} {\bibfnamefont {Y.~K.}\ \bibnamefont
  {Koh}}, \bibinfo {author} {\bibfnamefont {M.~H.}\ \bibnamefont {Bae}},
  \bibinfo {author} {\bibfnamefont {D.~G.}\ \bibnamefont {Cahill}}, \ and\
  \bibinfo {author} {\bibfnamefont {E.}~\bibnamefont {Pop}},\ }\href@noop {}
  {\bibfield  {journal} {\bibinfo  {journal} {Nano Lett.}\ }\textbf {\bibinfo
  {volume} {10}},\ \bibinfo {pages} {4363} (\bibinfo {year}
  {2010})}\BibitemShut {NoStop}%
\bibitem [{\citenamefont {Dorgan}\ \emph {et~al.}(2013)\citenamefont {Dorgan},
  \citenamefont {Behnam}, \citenamefont {Conley}, \citenamefont {Bolotin},\
  and\ \citenamefont {Pop}}]{Dorgan13}%
  \BibitemOpen
  \bibfield  {author} {\bibinfo {author} {\bibfnamefont {V.~E.}\ \bibnamefont
  {Dorgan}}, \bibinfo {author} {\bibfnamefont {A.}~\bibnamefont {Behnam}},
  \bibinfo {author} {\bibfnamefont {H.~J.}\ \bibnamefont {Conley}}, \bibinfo
  {author} {\bibfnamefont {K.~I.}\ \bibnamefont {Bolotin}}, \ and\ \bibinfo
  {author} {\bibfnamefont {E.}~\bibnamefont {Pop}},\ }\href@noop {} {\bibfield
  {journal} {\bibinfo  {journal} {Nano Lett.}\ }\textbf {\bibinfo {volume}
  {13}},\ \bibinfo {pages} {4581} (\bibinfo {year} {2013})}\BibitemShut
  {NoStop}%
\bibitem [{Rno()}]{Rnote}%
  \BibitemOpen
  \href@noop {} {\bibinfo  {journal} {As showed in Fig. 2(a)-inset, the
  $I-V_{B}$ characteristics at very low $V_{B}$ are precisely linear (no Joule
  heating). In which case $V_{B}/I = dV_{B}/dI$, and we use the slope to
  extract $R$ to avoid an error due to a very small (experimental) offset in
  $V_{B}$ (few 10s of micro-Volt). At higher bias, this small offset is
  negligible and we can safely use $R = V_{B}/I$. In Figure 2(c), $T_{e}$ is
  not constant versus $V_{B}$ due to Joule heating, thus $dV_{B}/dI$ also
  contains information about how quickly the temperature is changing with
  $V_{B}$, rather than only the temperature at one specific $V_{B}$ value.
  Since ($\Delta T$) is small, we find no significant quantitative difference
  in our results using either $dV_{B}/dI$ or $V_{B}/I$ to extract $T_{e}$, but
  the correct quantity which represents $T_{e}$ is $V_{B}/I$.}\ }\BibitemShut
  {NoStop}%
\bibitem [{\citenamefont {Du}\ \emph {et~al.}(2008)\citenamefont {Du},
  \citenamefont {Skachko}, \citenamefont {Barker},\ and\ \citenamefont
  {Andrei}}]{Du08}%
  \BibitemOpen
\bibfield  {journal} {  }\bibfield  {author} {\bibinfo {author} {\bibfnamefont
  {X.}~\bibnamefont {Du}}, \bibinfo {author} {\bibfnamefont {I.}~\bibnamefont
  {Skachko}}, \bibinfo {author} {\bibfnamefont {A.}~\bibnamefont {Barker}}, \
  and\ \bibinfo {author} {\bibfnamefont {E.~Y.}\ \bibnamefont {Andrei}},\
  }\href@noop {} {\bibfield  {journal} {\bibinfo  {journal} {Nature Nanotech.}\
  }\textbf {\bibinfo {volume} {3}},\ \bibinfo {pages} {491--495} (\bibinfo
  {year} {2008})}\BibitemShut {NoStop}%
\bibitem [{\citenamefont {Das~Sarma}\ \emph {et~al.}(2011)\citenamefont
  {Das~Sarma}, \citenamefont {Adam}, \citenamefont {Hwang},\ and\ \citenamefont
  {Rossi}}]{DasSarma11}%
  \BibitemOpen
  \bibfield  {author} {\bibinfo {author} {\bibfnamefont {S.}~\bibnamefont
  {Das~Sarma}}, \bibinfo {author} {\bibfnamefont {S.}~\bibnamefont {Adam}},
  \bibinfo {author} {\bibfnamefont {E.~H.}\ \bibnamefont {Hwang}}, \ and\
  \bibinfo {author} {\bibfnamefont {E.}~\bibnamefont {Rossi}},\ }\href@noop {}
  {\bibfield  {journal} {\bibinfo  {journal} {Rev. Mod. Phys.}\ }\textbf
  {\bibinfo {volume} {83}},\ \bibinfo {pages} {407} (\bibinfo {year}
  {2011})}\BibitemShut {NoStop}%
\bibitem [{\citenamefont {Li}\ and\ \citenamefont {Das~Sarma}(2013)}]{Li13}%
  \BibitemOpen
  \bibfield  {author} {\bibinfo {author} {\bibfnamefont {Q.}~\bibnamefont
  {Li}}\ and\ \bibinfo {author} {\bibfnamefont {S.}~\bibnamefont {Das~Sarma}},\
  }\href@noop {} {\bibfield  {journal} {\bibinfo  {journal} {Phys. Rev. B}\
  }\textbf {\bibinfo {volume} {87}},\ \bibinfo {pages} {085406} (\bibinfo
  {year} {2013})}\BibitemShut {NoStop}%
\bibitem [{VBn()}]{VBnote}%
  \BibitemOpen
  \href@noop {} {\bibinfo  {journal} {Using $n_{tot}(T)$ (SM section
  3)\cite{Dorgan10, Yigen13}, we define an effective chemical potential
  $\mu_{eff}(T) = \hbar v_{F}\sqrt{\pi n_{tot}(T)}$. For instance, for Sample A
  at $V_{G}=$ -5.3 V and $T$ = 100 K, $\mu_{eff}(100 K) =$ 49 meV. The various
  $V_{B}$ necessary to achieve $\Delta T \leq$ 10 K in Fig. 3 are always
  significantly smaller than $\mu_{eff}(T)$ and never larger than 27 mV. We
  only observe a change in the extracted $K_{e}$ values (in the doped-regime)
  when $\Delta T$ exceeds 20 K, and $V_{B} > \mu_{eff}(T)$.}\ }\BibitemShut
  {NoStop}%
\bibitem [{\citenamefont {Zuev}, \citenamefont {Chang},\ and\ \citenamefont
  {Kim}(2009)}]{Zuev09}%
  \BibitemOpen
\bibfield  {journal} {  }\bibfield  {author} {\bibinfo {author} {\bibfnamefont
  {Y.~M.}\ \bibnamefont {Zuev}}, \bibinfo {author} {\bibfnamefont
  {W.}~\bibnamefont {Chang}}, \ and\ \bibinfo {author} {\bibfnamefont
  {P.}~\bibnamefont {Kim}},\ }\href@noop {} {\bibfield  {journal} {\bibinfo
  {journal} {Phys. Rev. Lett.}\ }\textbf {\bibinfo {volume} {102}},\ \bibinfo
  {pages} {096807} (\bibinfo {year} {2009})}\BibitemShut {NoStop}%
\bibitem [{\citenamefont {Hwang}, \citenamefont {Rossi},\ and\ \citenamefont
  {Das~Sarma}(2009)}]{Hwang09}%
  \BibitemOpen
  \bibfield  {author} {\bibinfo {author} {\bibfnamefont {E.~H.}\ \bibnamefont
  {Hwang}}, \bibinfo {author} {\bibfnamefont {E.}~\bibnamefont {Rossi}}, \ and\
  \bibinfo {author} {\bibfnamefont {S.}~\bibnamefont {Das~Sarma}},\ }\href@noop
  {} {\bibfield  {journal} {\bibinfo  {journal} {Phys. Rev. B}\ }\textbf
  {\bibinfo {volume} {80}} (\bibinfo {year} {2009})}\BibitemShut {NoStop}%
\bibitem [{\citenamefont {Bian}\ \emph {et~al.}(2007)\citenamefont {Bian},
  \citenamefont {Zebarjadi}, \citenamefont {Singh}, \citenamefont {Ezzahri},
  \citenamefont {Shakouri}, \citenamefont {Zeng}, \citenamefont {Bahk},
  \citenamefont {Bowers}, \citenamefont {Zide},\ and\ \citenamefont
  {Gossard}}]{Bian07}%
  \BibitemOpen
  \bibfield  {author} {\bibinfo {author} {\bibfnamefont {Z.}~\bibnamefont
  {Bian}}, \bibinfo {author} {\bibfnamefont {M.}~\bibnamefont {Zebarjadi}},
  \bibinfo {author} {\bibfnamefont {R.}~\bibnamefont {Singh}}, \bibinfo
  {author} {\bibfnamefont {Y.}~\bibnamefont {Ezzahri}}, \bibinfo {author}
  {\bibfnamefont {A.}~\bibnamefont {Shakouri}}, \bibinfo {author}
  {\bibfnamefont {G.}~\bibnamefont {Zeng}}, \bibinfo {author} {\bibfnamefont
  {J.~H.}\ \bibnamefont {Bahk}}, \bibinfo {author} {\bibfnamefont {J.~E.}\
  \bibnamefont {Bowers}}, \bibinfo {author} {\bibfnamefont {J.~M.~O.}\
  \bibnamefont {Zide}}, \ and\ \bibinfo {author} {\bibfnamefont {A.~C.}\
  \bibnamefont {Gossard}},\ }\href@noop {} {\bibfield  {journal} {\bibinfo
  {journal} {Phys. Rev. B}\ }\textbf {\bibinfo {volume} {76}},\ \bibinfo
  {pages} {205311} (\bibinfo {year} {2007})}\BibitemShut {NoStop}%
\bibitem [{\citenamefont {Minnich}\ \emph {et~al.}(2009)\citenamefont
  {Minnich}, \citenamefont {Dresselhaus}, \citenamefont {Ren},\ and\
  \citenamefont {Chen}}]{Minnich09}%
  \BibitemOpen
  \bibfield  {author} {\bibinfo {author} {\bibfnamefont {A.~J.}\ \bibnamefont
  {Minnich}}, \bibinfo {author} {\bibfnamefont {M.~S.}\ \bibnamefont
  {Dresselhaus}}, \bibinfo {author} {\bibfnamefont {Z.~F.}\ \bibnamefont
  {Ren}}, \ and\ \bibinfo {author} {\bibfnamefont {G.}~\bibnamefont {Chen}},\
  }\href@noop {} {\bibfield  {journal} {\bibinfo  {journal} {Energy and
  Environmental Science}\ }\textbf {\bibinfo {volume} {2}},\ \bibinfo {pages}
  {466} (\bibinfo {year} {2009})}\BibitemShut {NoStop}%
\bibitem [{\citenamefont {Balandin}(2011)}]{Balandin11}%
  \BibitemOpen
  \bibfield  {author} {\bibinfo {author} {\bibfnamefont {A.~A.}\ \bibnamefont
  {Balandin}},\ }\href@noop {} {\bibfield  {journal} {\bibinfo  {journal}
  {Nature Mater.}\ }\textbf {\bibinfo {volume} {10}},\ \bibinfo {pages} {569}
  (\bibinfo {year} {2011})}\BibitemShut {NoStop}%
\bibitem [{\citenamefont {Pop}, \citenamefont {Varshney},\ and\ \citenamefont
  {Roy}(2012)}]{Pop12}%
  \BibitemOpen
  \bibfield  {author} {\bibinfo {author} {\bibfnamefont {E.}~\bibnamefont
  {Pop}}, \bibinfo {author} {\bibfnamefont {V.}~\bibnamefont {Varshney}}, \
  and\ \bibinfo {author} {\bibfnamefont {A.~K.}\ \bibnamefont {Roy}},\
  }\href@noop {} {\bibfield  {journal} {\bibinfo  {journal} {MRS Bulletin}\
  }\textbf {\bibinfo {volume} {37}},\ \bibinfo {pages} {1273} (\bibinfo {year}
  {2012})}\BibitemShut {NoStop}%
\end{thebibliography}
\end{document}